\documentclass{article}
\PassOptionsToPackage{numbers}{natbib}

\usepackage[preprint]{neurips_2024}

\usepackage[utf8]{inputenc} 
\usepackage[T1]{fontenc}    
\usepackage{hyperref}       
\usepackage{url}            
\usepackage{booktabs}       
\usepackage{amsfonts}       
\usepackage{nicefrac}       
\usepackage{microtype}      
\usepackage{xcolor}         
\usepackage{graphicx}
\usepackage{multirow}
\usepackage{wrapfig}
\usepackage{mathtools}

\definecolor{ao(english)}{rgb}{0.0, 0.5, 0.0}

\title{Item-Language Model for Conversational Recommendation}

\newcommand{\vvcomment}[1]{}

\author{%
    Li Yang \\
    Google Research\\
    \texttt{lyliyang@google.com} \\
    \And
    Anushya Subbiah \\
    Google Research \\
    \texttt{anushyas@google.com} \\
    \AND
    Hardik Patel \\
    Google \\
    \texttt{patelhardik@google.com} \\
    \And
    Judith~Yue Li \\
    Google Research \\
    \texttt{judithyueli@google.com} \\
    \And
    Yanwei Song \\
    Google \\
    \texttt{yanweisong@google.com} \\
    \And
    Reza Mirghaderi \\
    Google \\
    \texttt{rezam@google.com} \\
    \And
    Vikram Aggarwal \\
    Google Research \\
    \texttt{viki@google.com} \\
    \And
    Qifan Wang \\
    Meta AI \\
    \texttt{wqfcr@meta.com} \\
}

\begin{document}

\maketitle

\vvcomment{Top level todo:  empty.}

\begin{abstract}
  Large-language Models (LLMs) have been extremely successful at tasks like complex dialogue understanding, reasoning and coding due to their emergent abilities. These emergent abilities have been extended with multi-modality to include image, audio, and video capabilities. Recommender systems, on the other hand, have been critical for information seeking and item discovery needs. Recently, there have been attempts to apply LLMs for recommendations. One difficulty of current attempts is that the underlying LLM is usually not trained on the recommender system data, which largely contains user interaction signals and is often not publicly available. Another difficulty is user interaction signals often have a different pattern from natural language text, and it is currently unclear if the LLM training setup can learn more non-trivial knowledge from interaction signals compared with traditional recommender system methods. Finally, it is difficult to train multiple LLMs for different use-cases, and to retain the original language and reasoning abilities when learning from recommender system data. To address these three limitations, we propose an Item-Language Model (ILM), which is composed of an item encoder to produce text-aligned item representations that encode user interaction signals, and a frozen LLM that can understand those item representations with preserved pretrained knowledge. We conduct extensive experiments which demonstrate both the importance of the language-alignment and of user interaction knowledge in the item encoder.
\end{abstract}

\section{Introduction}

Large Language Models (LLMs), trained on web-scale data using a very large number of parameters, have shown remarkable emergent capabilities, such as in-context learning, reasoning, coding~\cite{gpt3, palm, palm2}. Recently, those abilities have been extended to multimodal domains, including image, audio, and video~\cite{gpt4, gemini, gemini15}. On various professional and academic benchmarks, those models achieve or surpass human-level performance~\cite{hosking2024human}. By contrast, the improvement in recommender systems have not derived similar breakthrough with improvements in LLM, even though recommender systems support a large volume of online user activity.  One reason is that users' interactions with recommender systems are typically not in natural language form. For example, a video recommender system typically recommends videos to users based on their implicit preferences from watch history and other query and candidate features, and users usually do not provide natural language preferences. In order to induce such natural language preferences, conversational recommendations have been proposed~\cite{Croft1987I3RAN,Belkin1995,Gker2000TheAP}, where the user can interact with the system explicitly using either natural-language tags or conversational natural language. With the development of techniques such as LLM tool-use~\cite{toolformer, qin2023toolllm}, Retrieval Augmented Generation~\cite{rag, gao2024retrievalaugmented}, and Agents~\cite{generative_agents,agents_review}, LLMs have demonstrated breakthroughs in many domains. It is possible that those technical advances and the resulting user habit changes will lead to conversational recommender systems to be more important in the future, thus it is worthwhile to develop methods to bring these benefits to conversational recommender systems.

\begin{figure}
  \centering
  \includegraphics[width=0.8\textwidth]{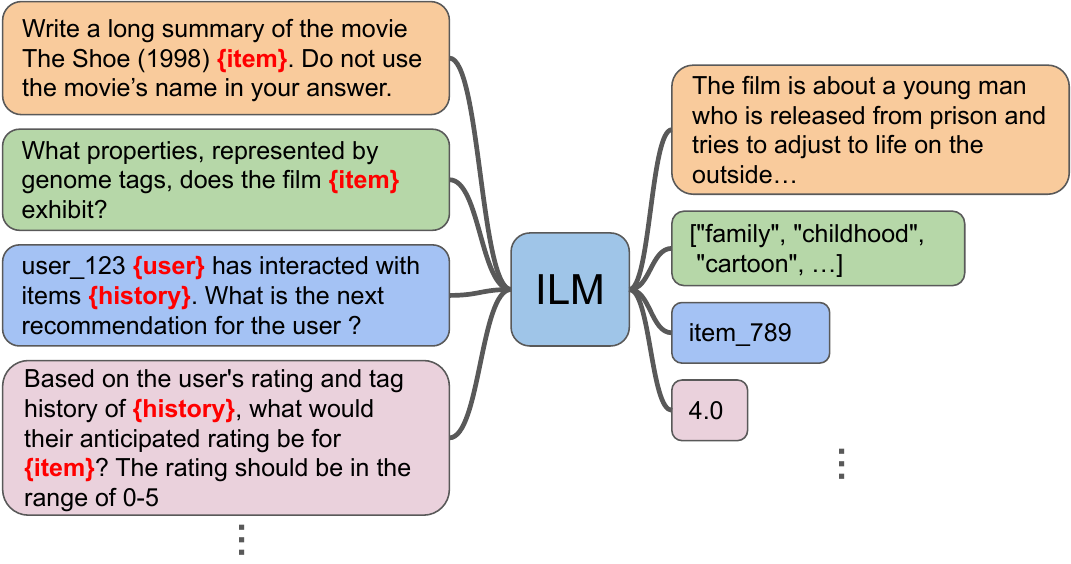}
  \caption{Conversational recommendation tasks using ILM. User and item collaborative filtering embeddings, marked by placeholders in the input, are interleaved with text embeddings and fed to the model. Where \{history\} is a sequence of items.}
  \label{ilm_model_tasks}
\end{figure}

However, using an LLM for conversational recommender systems has the following difficulties. (1)~Unlike multimodality, to the best of our knowledge no current LLM natively understands both user interaction signals and natural language. Current LLMs are trained on natural language data or natural language aligned contents, e.g. image with description, audio with transcription, etc. User interaction signals exist in the form of item co-watch, for example, if many users have viewed both $v_1$ and $v_2$, then a user who likes $v_1$ may also like $v_2$. Those signals often have a complicated nature, e.g. it is hard to explain why people tend to watch $v_1$ and $v_2$ together, so there is often a lack of natural language descriptions of such signals. Due to this, existing approaches using In-Context Learning~\cite{gpt3} by prompting LLMs for recommendation tasks purely using natural language struggle to achieve superior performance~\cite{Dai2023,gao2023chatrec,kang2023llms,Zhang2021,wang-etal-2023-chatgpt}. Other approaches finetuning LLMs on user interaction signals~\cite{Bao_2023, kang2023llms} show improvement upon purely prompting based methods. However, after such finetuning, the LLM's original abilities are often forgotten or hidden, and new privacy concerns may arise if user data are not handled correctly~\cite{hu2024exact}. (2)~Although it is possible to annotate an item with a text format context of all related items/users, or a user with a text format context of all related items, by doing so, the context for the user/item could be very long, which will dramatically increase the inference cost. (3)~Finally, another way to incorporate behavioral data is to first compute item and user collaborative filtering embeddings using matrix factorization algorithms, then feed those embeddings to LLM through a mapping module~\cite{tennenholtz2024demystifying,zhang2023collm}. This approach introduces a modality gap between collaborative embeddings and LLM pretrain token embeddings, so the LLM still needs to be finetuned to close this modality gap.

\begin{figure}[t]
  \centering
  \includegraphics[width=\textwidth]{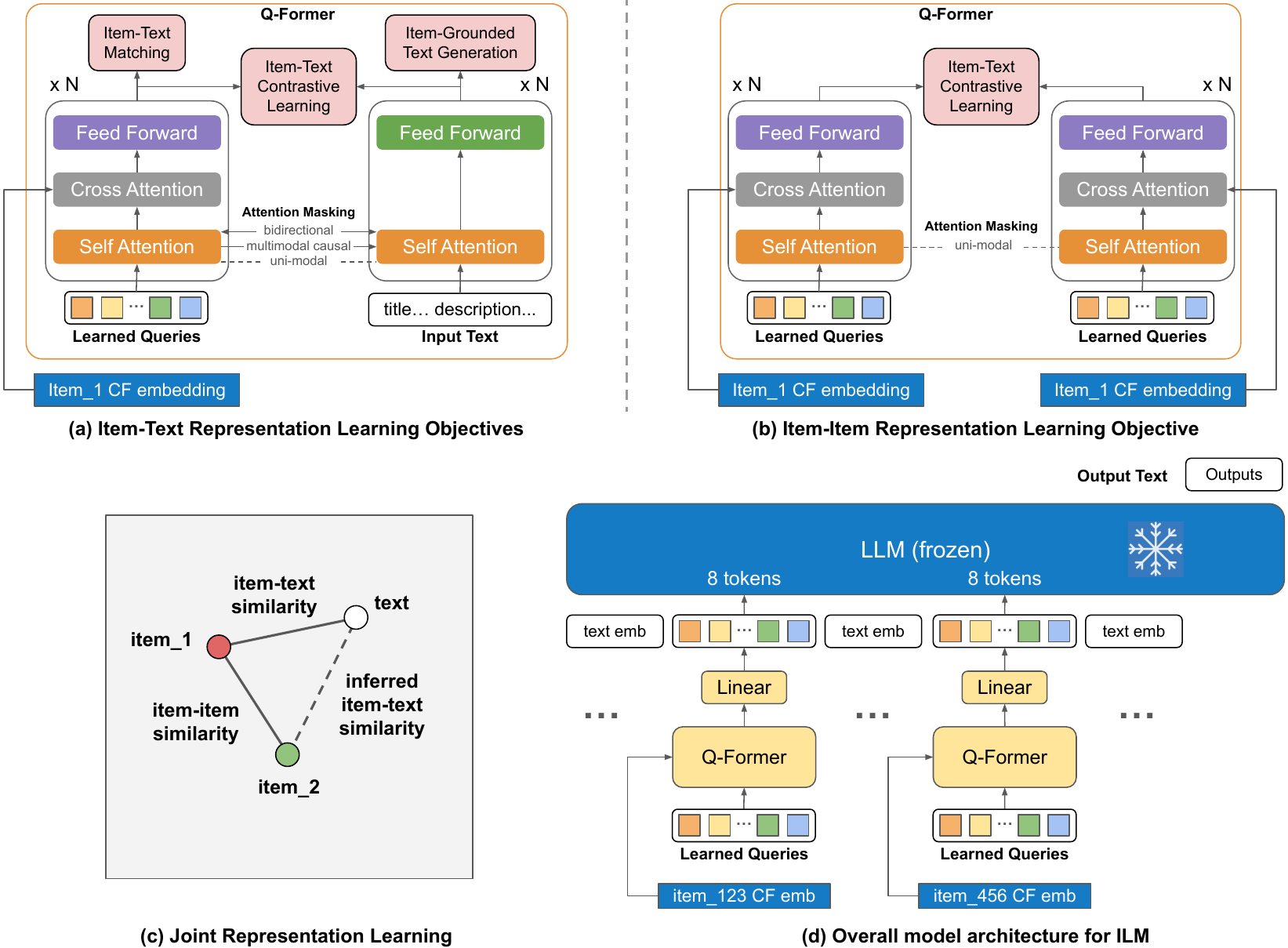}
  \caption{Overall model architecture for ILM. {\bf (a)} The original item-text contrastive, item-grounded text generation and item-text matching losses used in BLIP-2~\cite{blip2} in Q-Former phase 1 training. {\bf (b)} The new item-item contrastive loss we introduced in Q-Former phase 1 training. For user-item contrastive learning, we simply replace item collaborative filtering (CF) embedding with user CF embedding. {\bf (c)} A schematic of how item-item contrastive learning can improve text-aligned item representations. {\bf (d)} The ILM phase 2 training by integrating the Q-Former to a frozen LLM.}
  \label{ilm_model_overall}
\end{figure}

Towards solving the above difficulties, we propose Item-Language Model (ILM, hereafter) for conversational recommendation tasks, Figure~\ref{ilm_model_tasks}. ILM is a two-phase framework containing an item-language representation learning phase and an item-language model training phase. The overall model architecture is shown in Figure~\ref{ilm_model_overall}(d). Motivated by the BLIP-2~\cite{blip2} work of bridging the modality gap with a lightweight Querying Transformer (Q-Former), we adopt Q-Former to generate item-language aligned representations. In Phase 1, we pretrain a Q-Former encoder following the BLIP-2 approach. In addition to the original item-text objectives, we also introduce an item-item contrastive objective, which plays a regularization role and also encodes co-watch information in the resulting item-language representation. The inputs to the item encoder are collaborative filtering embeddings. Throughout the paper, we treat the user as a special item. In Phase 2, we integrate the Q-Former into a pretrained LLM through a linear projection adaptor layer, and finetune the ILM on conversational recommendation tasks in multitasking fashion. During finetuning, only the Q-Former and the adaptor parameters are updated, the pretrained LLM is kept frozen to preserve its pretrained abilities. The main contributions of this work are summarized as follows:

\begin{enumerate}
  \item We propose ILM, a novel framework for using a Q-Former item encoder to produce item-language aligned representations from collaborative filtering embeddings, then integrate into a frozen LLM for conversation recommendation tasks with interleaved item-text inputs.
  \item We conduct extensive benchmarks and ablation studies on various conversation recommendation tasks to show that our ILM approach consistently outperforms existing approaches of integrating item representations into LLMs across all tasks.
\end{enumerate}

\paragraph{Paper Overview.} In Section~\ref{sec_related_work}, we provide a brief literature survey of LLM for recommender systems, how item representations are used there and multimodality LLM methods. In Section~\ref{sec_method}, we present the details of the proposed ILM approach, including model architecture and training. We present our experiment results in Section~\ref{sec_experiments} and ablation studies Section~\ref{sec_ablation_study}. We conclude in Section~\ref{sec_conclusion}.

\section{Related Work}
\label{sec_related_work}

\paragraph{LLM for recommendation}
With LLMs showing remarkable emergent abilities and surpassing human-level performance across various domains, there have been explorations to apply LLMs to recommender systems~\cite{Mohanty2024,wu2023survey}. In-Context Learning methods have been used as a straightforward way for this purpose~\cite{Dai2023,gao2023chatrec,kang2023llms,Zhang2021,wang-etal-2023-chatgpt}, which rely on LLMs' world knowledge. However, since user interaction data in recommender systems is largely not available during LLM pretraining, purely text-prompting based methods show suboptimal performance. Another line of work is by finetuning a language model on user interaction data. P5~\cite{p5, xu2023openp5} pretrains a unified language model for many different recommendation tasks by converting them into a common natural language sequence format, where user and item ids are represented as text strings. TALLRec~\cite{Bao_2023} finetunes a LLM using LoRA~\cite{hu2022lora} on user rating data, and the model outputs a binary label.

\paragraph{Item representations in LLM4Rec}
Efficiently representing users and items in recommender systems is a rich field with years of work of traditional techniques such as Matrix Factorization~\cite{Koren2009mf,MF}. When applying LLM to recommender systems, users and items are key objects, and it is critical for LLM to be able to understand them. Using text representation, such as the title of an item is a straightforward way. However, one issue is text representation of an item may not be informative enough. For example, it is quite common that a video title is unrelated to the video content, and different content can have a single title. We can always include more text features of an item, such as description, author and other content features, but this may introduce irrelevant information, which may confuse the model or make the method inefficient due to very long input. Quantized id representations as item representation has been proposed in the methods of random indexing~\cite{anderson2020superbloom,xu2023openp5}, sequential or collaborative indexing~\cite{xu2023openp5,hua2023index} and semantic ids in the context of generative retrieval~\cite{singh2023better,generative_retrieval_1,generative_retrieval_2}. In addition to using quantized ids, ELM~\cite{tennenholtz2024demystifying} demystifies the input embedding spaces by feeding semantic embeddings to LLM, and CoLLM~\cite{zhang2023collm} enhances recommendation performance on rating prediction tasks by feeding user and item collaborative filtering embeddings to LLM. Recently, USER-LLM~\cite{ning2024userllm} contextualizes LLM with user history embeddings by integrating user embeddings to LLM through perceiver~\cite{perceiver,flamingo}, projection, and cross-attention modules, where the LLM can be frozen to preserve the original ability.

\paragraph{Multimodal LLM}
Our approach is closely related to a series of multimodal foundation models trained with both generative and contrastive learning objectives, including BLIP-2~\citep{blip2}, CoCa~\citep{yu2022coca}, and MaMMUT~\citep{kuo2023mammut}, which achieve state-of-art performance on vision-language tasks. In our case the pretrained LLM is kept frozen, requiring a pretrained adaptor with enough representation capacity like Q-Former~\citep{blip2} used in BLIP-2 and GILLMapper~\citep{koh2023generating}. Compared to CogVLM~\citep{wang2024cogvlm}, which introduces a visual module to modify the frozen LLM directly, our approach is considered as late fusion. Our two-phase workflow mostly resemble BLIP-2's two-phase workflow, including pretraining an adapter in phase one and task fine-tuning in phase two. Compared to general purpose vision-language model~\cite{chen2023pali, yu2022parti}, our approach introduces a novel item-item contrastive loss and deals with interleaved item and text inputs~\cite{driess2023palme}. Our approach of keeping LLM frozen and only finetune the item encoder and adaptor is related to Frozen~\cite{frozen}, which exhibits great multimodal few-shot learning capability.


\section{Method}
\label{sec_method}
\subsection{Model Architecture}
For conversational recommendation tasks, the input and output of the model are typically sequences of interleaved items and text. In addition, for recommendation tasks, the model's ability of understanding items is critical. In this work, we focus on the input item representation with the overall ILM model architecture shown in Figure~\ref{ilm_model_overall}(d). Collaborative filtering information is widely believed to be very important for recommender system performance, and forms the basis of many existing recommender systems. Existing work includes feeding item and user's collaborative filtering embeddings into LLM through a two layer MLP~\cite{zhang2023collm}. As can be seen in Figure~\ref{ilm_model_overall} we adopt the Q-Former~\cite{blip2} architecture to encode items, combined with an item-text alignment pretraining phase to allow the Q-Former to be able to produce text-aligned item representations that also encode collaborative filtering information. We show significant performance gains of this approach on various conversational recommendation tasks in Section~\ref{sec_experiments}. The two phases of our proposed ILM model training are discussed in Section~\ref{phase1} and Section~\ref{phase2}, respectively.

\subsection{Item-Language Representation Learning}
\label{phase1}
In phase 1, we pretrain the Q-Former item encoder to ensure that it can generate text-aligned item representations given item collaborative filtering embeddings. We adopt the original 3 tasks of item-text contrastive learning, item-text generation and item-text matching in BLIP-2~\cite{blip2}, Figure~\ref{ilm_model_overall}(a). For item-grounded text generation, we use auto-regressive loss on top of the text tower. For item-text matching, we use a binary cross-entropy loss on the CLS token output from the text tower. For item-text contrastive learning, we compute the loss as follows. Given a positive item-text pair, we first compute output representations $[h_1, h_2, ..., h_N]$ of the $N$ learnable queries $[q_1, q_2, ..., q_N]$ from the query tower $f_q$ and use the CLS output representation $h_{\rm cls}$ from the text tower $f_t$ as the text representation:
\begin{equation}
\label{eq_item_text_contrastive_1}
    \begin{split}
        [h_1, h_2, ..., h_N] &= f_q([q_1, q_2, ..., q_N], e) \\
        h_{\rm cls} &= f_t([t_{\rm cls}, t_1, ..., t_L]), \\
    \end{split}
\end{equation}
where $e$ is the item input embedding and $[t_1, ..., t_L]$ are text tokens. We select the closest query representation to $h_{\rm cls}$ as the item representation:
\begin{equation}
\label{eq_item_text_contrastive_2}
    \begin{split}
        i = {\rm argmax}_j\;\; &{\rm cosine\_similarity}(h_j, h_{\rm cls}) \\
        &h_{\rm item}=h_i
    \end{split}
\end{equation}
Then we compute the contrastive loss between $h_{\rm item}$ and $h_{\rm cls}$ using in-batch negative sampling following~\cite{clip}. The item-text pair data we used is processed from item descriptions and tags, more details can be found in Section~\ref{sec_datasets}. In addition to item-text alignment tasks, we also introduce a novel item-item contrastive learning loss to prevent the model from overfitting on item-text data when text labels are sparse, Figure~\ref{ilm_model_overall}(b). The item-item contrastive loss is computed in a similar way by extending the item-text contrastive loss, Equation~\ref{eq_item_text_contrastive_1} and Equation~\ref{eq_item_text_contrastive_2}. Given a positive item-item pair, we first compute output representations of the N learnable queries for both items:
\begin{equation}
\label{eq_item_item_contrastive_1}
    \begin{split}
        [h^{[1]}_1, h^{[1]}_2, ..., h^{[1]}_N] &= f_q([q_1, q_2, ..., q_N], e^{[1]}) \\
        [h^{[2]}_1, h^{[2]}_2, ..., h^{[2]}_N] &= f_q([q_1, q_2, ..., q_N], e^{[2]}), \\
    \end{split}
\end{equation}
where $e^{[1]}$ and $e^{[2]}$ are embeddings for item1 and item2, respectively. Note that we use a single set of learnable queries for all items. We select the pair of closest query representations as the representations for the two items:
\begin{equation}
\label{eq_item_item_contrastive_2}
    \begin{split}
        k,l = {\rm argmax}_{i,j}\;\; &{\rm cosine\_similarity}(h_i, h_j) \\
        h_{\rm item1}&=h_k;\;\;h_{\rm item2}=h_l
    \end{split}
\end{equation}
Then we compute the contrastive loss between $h_{\rm item1}$ and $h_{\rm item2}$ in the same way as item-text contrastive loss. Introducing the item-item contrastive loss can provide an additional benefit, namely that the Q-Former output is a better representation. This representation not only encodes item-text similarity knowledge but also encodes item-item similarity knowledge, so that for items without any text labels, it can encode certain text information indirectly, as depicted in Figure~\ref{ilm_model_overall}(c). Due to the in-batch negative sampling of contrastive loss, we cannot mix item-text and item-item examples in the same batch. To address this, we train Q-Former on item-text and item-item batches in an alternating manner. More results on the effects of the item-item contrastive loss are shown in Section~\ref{sec_item_item_effects}.

\subsection{Item-Language Model Training}
\label{phase2}
In phase 2, we integrate the Q-Former item encoder with a pretrained LLM. The LLM can be pretrained for any purpose, which could be a general purpose human instruction followed LLM~\cite{gpt4,gemini,gemini15} or a LLM trained on recommender system data that supports generative retrieval of candidate items~\cite{generative_retrieval_1,generative_retrieval_2}. We present experimental results for the former in Section~\ref{elm_results} and the latter in Section~\ref{openp5_results}. During training, we freeze the LLM parameters and only tune the parameters in the Q-Former encoder and the linear projection layer. The linear projection layer maps the Q-Former output dimension to the LLM input dimension as shown in Figure~\ref{ilm_model_overall}(d).

\section{Experiments}
\label{sec_experiments}
\subsection{Datasets}
\label{sec_datasets}

A conversational recommendation can consist of multiple sub-tasks, including user preference elicitation, recommendation, explanation, and item information search~\cite{feng2023large}. To benchmark our proposed ILM approach on all these tasks, we select the Embedding Language Model (ELM) dataset~\cite{tennenholtz2024demystifying} and the OpenP5 benchmark. The ELM dataset contains 24 tasks which covers user preference elicitation, explanation, and item information search, etc. The OpenP5 dataset contains the sequential recommendation task and the straightforward recommendation task.

\paragraph{ELM 24 Tasks}
The ELM 24 tasks~\cite{tennenholtz2024demystifying} are created from the MovieLens 25M dataset, including single movie tasks, e.g. summarizing a movie, and movie pair tasks, e.g. comparing characteristics of movies. The training targets are generated by prompting the PaLM 2-L model with a movie's title and task-specific information. For training inputs, the same task-specific prompts are used, but with the movie title replaced by the movie embedding. We use the behavioral embedding in the ELM dataset, which are trained on user ratings in the MovieLens 25M dataset with Matrix Factorization computed using Weighted Alternating Least Squares (WALS)~\cite{wals}.


\paragraph{OpenP5}
OpenP5~\cite{xu2023openp5,hua2023index} is a dataset for LLM-based Recommendation development, finetuning, and evaluation. It provides 10 popular preprocessed public datasets, and each dataset contains two kinds of tasks: Sequential Recommendation and Straightforward Recommendation. We select the MovieLens-1M, Beauty, and Clothing datasets with random indexing item representation for our benchmarks, which can be thought of as a simple generative-retrieval setup. The training target for each example is the ground truth item id. For training inputs, we append each item's random indexing id with its behavioral embedding, which are computed using the iALS matrix factorization algorithm~\cite{MF} on the user sequence training set. We use the provided train, development, and test split in the OpenP5 dataset, which uses the last item in the user sequence for testing and the second from the last item in the user sequence for development.

\subsection{Evaluation Metrics}
For ELM 24 tasks, we report the log perplexity and Semantic Consistency (SC)~\cite{tennenholtz2024demystifying} on the test set. For SC, we use the cosine similarity of semantic embeddings of the original and decode targets from the Sentence-T5 11B model~\cite{st5}. For OpenP5 tasks, we report top-k Hit Rate (HR@K) and Normalized Discounted Cumulative Gain (NDCG@K) with K = 5, 10 to evaluate the recommendation performance. To compute those metrics, we use beam search to generate 10 outputs for each example, and remove invalid outputs that do not match the regular expression "\texttt{.*item\_(\textbackslash d+)\$}".

\begin{table}
  \caption{SC metrics on the ELM 24 tasks using item behavioral embedding. We define SC as the semantic embedding cosine similarity between the decoded text and original text. We adopt the Sentence-T5 11B model~\cite{st5} for computing semantic embeddings.}
  \label{elm_results_table}
  \centering
  \begin{tabular}{lcccccccc}
\toprule
\multirow{2}{*}{Tasks} & \multicolumn{2}{c}{MLP} & \multicolumn{2}{c}{ILM-rand} & \multicolumn{2}{c}{ILM} \\
\cmidrule(r){2-3} \cmidrule(r){4-5} \cmidrule(r){6-7}
& SC(\%) & Log pplx & SC(\%) & Log pplx & SC(\%) & Log pplx \\
\midrule
summary              & 71.68 & 0.6549 & 72.69 & 0.6257 & 73.88 & 0.6223 \\
positive review      & 76.42 & 0.6114 & 77.84 & 0.5708 & 78.81 & 0.5669 \\
neutral review       & 74.61 & 0.6167 & 75.98 & 0.5770 & 77.94 & 0.5690 \\
five pos char.       & 84.96 & 0.6778 & 85.74 & 0.5939 & 86.92 & 0.5850 \\
five neg char.       & 90.77 & 0.4953 & 92.46 & 0.4009 & 92.24 & 0.3874 \\
long description     & 71.09 & 0.6921 & 71.98 & 0.6629 & 73.39 & 0.6596 \\
funnier              & 69.56 & 0.5744 & 68.79 & 0.5401 & 69.62 & 0.5378 \\
sadder               & 70.57 & 0.5470 & 71.67 & 0.5163 & 72.31 & 0.5130 \\
scarier              & 71.55 & 0.5578 & 72.07 & 0.5258 & 72.90 & 0.5230 \\
improve              & 76.37 & 0.5607 & 77.11 & 0.5111 & 78.15 & 0.5073 \\
movie to viewer      & 76.93 & 0.6801 & 79.33 & 0.6041 & 80.70 & 0.5920 \\
pitch                & 78.95 & 0.5874 & 83.26 & 0.5165 & 84.28 & 0.5142 \\
criticize            & 76.62 & 0.6073 & 78.13 & 0.5514 & 79.63 & 0.5423 \\
convince1            & 77.16 & 0.6859 & 78.55 & 0.6196 & 79.48 & 0.6106 \\
convince2            & 78.48 & 0.8748 & 78.98 & 0.7842 & 80.53 & 0.7688 \\
convince3            & 78.02 & 0.9860 & 79.01 & 0.8955 & 80.33 & 0.8761 \\
dissuade1            & 77.90 & 0.6998 & 79.00 & 0.6107 & 79.49 & 0.6002 \\
dissuade2            & 80.97 & 0.8523 & 83.71 & 0.7296 & 84.15 & 0.7159 \\
\midrule
similarities         & 80.32 & 0.4993 & 83.24 & 0.4082 & 85.33 & 0.3828 \\
interpolation        & 71.43 & 0.5627 & 72.59 & 0.5332 & 73.70 & 0.5263 \\
why like nn          & 78.47 & 0.7494 & 79.59 & 0.6536 & 80.96 & 0.6343 \\
diff than nn         & 86.56 & 0.6127 & 88.60 & 0.5238 & 89.30 & 0.4923 \\
common with nn       & 80.76 & 0.7008 & 83.70 & 0.5466 & 84.62 & 0.5234 \\
\midrule
all                  & 77.48 & 0.6520 & 78.96 & 0.5837 & 80.01 & 0.5730 \\
& - & - & \color{ao(english)}{+1.91\%} & \color{ao(english)}{-10.47\%} & \color{ao(english)}{+3.27\%} & \color{ao(english)}{-12.12\%} \\
\bottomrule
  \end{tabular}
\end{table}

\subsection{Results on ELM 24 tasks}
\label{elm_results}
We show the results of our ILM approach on ELM 24 tasks in Table~\ref{elm_results_table}. We adopt the CoLLM~\cite{zhang2023collm} approach as a baseline, where a two layer MLP with intermediate size 10 times the input embedding size is used to map the input behavioral embedding to LLM token embedding space. For ILM, we use a Q-Former encoder with 8 transformer layers. For phase 1 training, we pair the item with a concatenation of the prompt to generate the original target, i.e. title(s) and task specific information, and the original target as the item-text pair data. For the ELM 24 tasks benchmark, we only train the Q-Former using item-text data to avoid data leakage from item-item data. For all models, we use PaLM 2-S~\cite{palm2} as the LLM backbone. We train the models for 100k steps using a batch size 32 and learning rate $5 \times 10^{-4}$ with a cosine decay.  We also add an {\bf ILM-rand} benchmark, which is an ILM model with a randomly initialized Q-Former encoder. As can be seen, our approach consistently outperforms the MLP baseline across all tasks. {\bf ILM} also outperforms {\bf ILM-rand} by a noticeable margin, which suggests the importance of the item-language representation learning phase. We report mean and standard error of our ILM results in Table~\ref{elm_mean_sem}.



\subsection{Results on OpenP5 tasks}
\label{openp5_results}
In the OpenP5 dataset, each task contains 10 prompt templates used for training, and 1 prompt template used for unseen testing. In Table~\ref{openp5_results}, we show seen and unseen test results on OpenP5 MovieLens-1M, Beauty, and Clothing tasks. For {\bf ILM}, we use a Q-Former encoder with 8 transformer layers.

For phase 1 training, we generate the item-text pair data by extracting item metadata from (1)~movie title and genres from the original ML1M dataset~\cite{ml1m_dataset} for the ML1M task (2)~product metadata including title, description, features, brand, etc. from the original Amazon Review 2014 Metadata~\cite{amazon_reviews_2014_1,amazon_reviews_2014_2} for Beauty and Clothing tasks. Since the inputs of OpenP5 tasks contain both user id and item id, we generate user-item pair data using the training set of OpenP5's user sequence data, and conduct user-item contrastive learning to phase 1 training.

For phase 2 training, we use an 8 layer transformer model as the LLM backbone, and pretrain the backbone on the purely text format OpenP5 data using random item indexing to enable the model generative retrieval ability. We believe our ILM approach can be integrated with any other kind of item token id based encoding such as sequential indexing and collaborative indexing in the OpenP5 dataset as well as other more advanced semantic id based  methods~\cite{generative_retrieval_1,generative_retrieval_2}. Training hyperparameters can be found in Section~\ref{sec_hparams}. We show the statistics of phase 1 and phase 2 data in Table~\ref{openp5_data_stats}.

We use the following baselines for the OpenP5 benchmark
\begin{itemize}
  \item {\bf OpenP5-R} stands for the OpenP5 random indexing method, i.e. using the backbone model directly without any embedding inputs.
  \item {\bf MLP} stands for using a two layer MLP with hidden size 10 times the input embedding size to map the input behavioral embedding to LLM token embedding space following CoLLM~\cite{zhang2023collm}. However, during the training, we do not finetune the LLM backbone in order to preserve the pretrained abilities.
\end{itemize}
We also add a {\bf ILM-rand} baseline, which is an ILM model with random initialized Q-Former encoder. For each dataset, we select the checkpoint with the best NDCG@10 metric on the development set. For all models, we use the same 8 layer transformer model as the LLM backbone. As can be seen in Table~\ref{openp5_results_table}, our method consistently outperforms other baselines across all datasets. {\bf ILM} also outperforms {\bf ILM-rand}, suggesting the importance of the item-language representation learning phase. We report mean and standard error of our ILM results in Table~\ref{openp5_mean_sem}.

\begin{table}[t!]
  \caption{OpenP5 phase 1 and phase 2 dataset statistics.}
  \label{openp5_data_stats}
  \centering
  \begin{tabular}{lrrrrrrrr}
\toprule
\multirow{2}{*}{Datasets} & \multicolumn{3}{c}{Phase 1} & \multicolumn{2}{c}{Phase 2} & & \\
\cmidrule(r){2-4} \cmidrule(r){5-6}
 & Item-text & Item-item & User-item & Train & Test & \# Users & \# Items \\
\midrule
ML1M       &     3079 &   479664 &   888696 & 19629820 & 12080 & 6040   & 3416\\
Beauty     &    10879 &   103268 &   138521 & 2628260  & 44726 & 22363  & 12101 \\
Clothing   &    20750 &   142427 &   180128 & 3210280  & 78774 & 39387  & 23033 \\
\bottomrule
  \end{tabular}
\end{table}

\begin{table}[t!]
  \caption{Results on OpenP5 tasks using item behavioral embedding.}
  \label{openp5_results_table}
  \centering
  \resizebox{\textwidth}{!}{
  \begin{tabular}{ccccccccccccc}
\toprule
\multirow{2}{*}{Methods} & \multicolumn{4}{c}{ML1M} & \multicolumn{4}{c}{Beauty} & \multicolumn{4}{c}{Clothing} \\
\cmidrule(r){2-5} \cmidrule(r){6-9} \cmidrule(r){10-13}
& HR@5 & NDCG@5 & HR@10 & NDCG@10 & HR@5 & NDCG@5 & HR@10 & NDCG@10 & HR@5 & NDCG@5 & HR@10 & NDCG@10 \\
\midrule
OpenP5-R(seen)            & {\rm 0.0688 } & {\rm  0.0455 } & {\rm  0.1033 } & {\rm  0.0566} & {\rm  0.0208 } & {\rm  0.0162 } & {\rm  0.0254 } & {\rm  0.0177} & {\rm  0.0015 } & {\rm  0.0010 } & {\rm  0.0030 } & {\rm  0.0015 } \\
MLP(seen)                  & {\rm  0.0692 } & {\rm  0.0459 } & {\rm  0.1041 } & {\rm  0.0572} & {\rm  0.0199 } & {\rm  0.0150 } & {\rm  0.0255 } & {\rm  0.0168} & {\rm  0.0015 } & {\rm  0.0010 } & {\rm  0.0022 } & {\rm  0.0012 } \\
ILM-rand(seen)             & {\rm  0.0723 } & {\rm  0.0478 } & {\bf  0.1084 } & {\rm  0.0594} & {\rm  0.0206 } & {\rm  0.0154 } & {\rm  0.0259 } & {\rm  0.0171} & {\rm  0.0026 } & {\rm  0.0017 } & {\rm  0.0038 } & {\rm  0.0021 }\\
ILM(seen)                  & {\bf  0.0724 } & {\bf  0.0485 } & {\rm  0.1064 } & {\bf  0.0595} & {\bf  0.0213 } & {\bf  0.0164 } & {\bf  0.0270 } & {\bf  0.0182} & {\bf  0.0041 } & {\bf  0.0025 } & {\bf  0.0065 } & {\bf  0.0033 }\\
\midrule
OpenP5-R(unseen)           & {\rm  0.0696 } & {\rm  0.0449 } & {\rm  0.1041 } & {\rm  0.0560} & {\rm  0.0206 } & {\rm  0.0161 } & {\rm  0.0253 } & {\rm  0.0176} & {\rm  0.0016 } & {\rm  0.0010 } & {\rm  0.0034 } & {\rm  0.0016 }\\
MLP(unseen)                & {\rm  0.0716 } & {\rm  0.0470 } & {\rm  0.1045 } & {\rm  0.0576} & {\rm  0.0203 } & {\rm  0.0151 } & {\rm  0.0255 } & {\rm  0.0168} & {\rm  0.0016 } & {\rm  0.0011 } & {\rm  0.0026 } & {\rm  0.0014 }\\
ILM-rand(unseen)           & {\rm  0.0710 } & {\rm  0.0477 } & {\rm  0.1033 } & {\rm  0.0581} & {\rm  0.0206 } & {\rm  0.0157 } & {\rm  0.0263 } & {\rm  0.0175} & {\rm  0.0023 } & {\rm  0.0015 } & {\rm  0.0037 } & {\rm  0.0019 }\\
ILM(unseen)                & {\bf  0.0717 } & {\bf  0.0481 } & {\bf  0.1086 } & {\bf  0.0600} & {\bf  0.0213 } & {\bf  0.0162 } & {\bf  0.0269 } & {\bf  0.0181} & {\bf  0.0038 } & {\bf  0.0024 } & {\bf  0.0062 } & {\bf  0.0032 } \\
\bottomrule
  \end{tabular}
  }
\end{table}

\section{Ablation Study}
\label{sec_ablation_study}
\subsection{Effects of Q-Former Phase 1 Training.}
\label{sec_item_item_effects}
As shown in Table~\ref{elm_results_table} and Table~\ref{openp5_results_table}, {\bf ILM} consistently outperforms {\bf ILM-rand} by a noticeable margin across all metrics on all benchmarks, which suggests the importance of the Q-Former phase 1 training. For the OpenP5 dataset, we experiment with different combinations of phase 1 training losses: (1) Only using item-text losses ({\bf  ILM-IT}). (2) Combine item-text losses with an item-item contrastive loss ({\bf ILM-IT-II}). (3) Combine item-text losses with an user-item contrastive loss ({\bf ILM-IT-UI}). We generate item-item pair data for (2) as follows. For each user, we treat two consecutive items in the history sequence as a positive pair, then we perform deduplication to get all unique pairs as the item-item pair data. The number of pairs generated are shown in Table~\ref{openp5_data_stats}.

The results for the above models are shown in Table~\ref{openp5_phase1_loss_ablate}. We observe that for ML1M dataset, introducing user-item or item-item contrastive losses can in general lead to performance gains, while for Beauty and Clothing there are no obvious gains. We hypothesize this is due to ML1M's item-text pair data being much scarcer and user interactions are much richer than the other two datasets. As can be seen in Table~\ref{openp5_data_stats}, comparing with other datasets, the ML1M dataset contains much fewer users and items, but much more user-item interactions. This supports our hypothesis, and suggests exploring user-interaction signals in the phase 1 representation learning can be beneficial for datasets like ML1M. To demonstrate the regularization effects of the item-item and user-item contrastive losses, we showed the phase 1 final train and eval item-grounded text generation losses in Figure~\ref{openp5_phase1_loss_values}. We can observe that adding item-item or user-item contrastive losses in phase 1 indeed can help to reduce the eval loss and close the train-eval gap.

\begin{table}[t!]
  \caption{Effects of phase 1 item-item and user-item contrastive losses on OpenP5 benchmarks.}
  \label{openp5_phase1_loss_ablate}
  \centering
  \resizebox{\textwidth}{!}{
  \begin{tabular}{ccccccccccccc}
\toprule
\multirow{2}{*}{Methods} & \multicolumn{4}{c}{ML1M} & \multicolumn{4}{c}{Beauty} & \multicolumn{4}{c}{Clothing} \\
\cmidrule(r){2-5} \cmidrule(r){6-9} \cmidrule(r){10-13}
& HR@5 & NDCG@5 & HR@10 & NDCG@10 & HR@5 & NDCG@5 & HR@10 & NDCG@10 & HR@5 & NDCG@5 & HR@10 & NDCG@10 \\
\midrule
ILM-IT(seen)              & {\rm 0.0719} & {\rm 0.0474} & {\rm 0.1088} & {\rm 0.0594} & {\rm 0.0212} & {\rm 0.0160} & {\rm 0.0262} & {\rm 0.0177} & {\bf 0.0044} & {\bf 0.0029} & {\rm 0.0061} & {\bf 0.0035} \\
ILM-IT-II(seen)           & {\rm 0.0712} & {\rm 0.0479} & {\bf 0.1093} & {\bf 0.0602} & {\rm 0.0210} & {\rm 0.0160} & {\rm 0.0261} & {\rm 0.0177} & {\rm 0.0040} & {\rm 0.0027} & {\rm 0.0060} & {\rm 0.0033} \\
ILM-IT-UI(seen)           & {\bf 0.0724} & {\bf 0.0485} & {\rm 0.1064} & {\rm 0.0595} & {\bf 0.0213} & {\bf 0.0164} & {\bf 0.0270} & {\bf 0.0182} & {\rm 0.0041} & {\rm 0.0025} & {\bf 0.0065} & {\rm 0.0033} \\
\midrule
ILM-IT(unseen)            & {\rm 0.0700} & {\rm 0.0470} & {\rm 0.1071} & {\rm 0.0589}& {\bf 0.0218} & {\bf 0.0163} & {\bf 0.0275} & {\bf 0.0182} & {\bf 0.0039} & {\bf 0.0025} & {\rm 0.0056} & {\rm 0.0031} \\
ILM-IT-II(unseen)         & {\rm 0.0701} & {\rm 0.0472} & {\rm 0.1078} & {\rm 0.0594}& {\rm 0.0216} & {\rm 0.0162} & {\rm 0.0269} & {\rm 0.0180} & {\rm 0.0037} & {\rm 0.0024} & {\rm 0.0054} & {\rm 0.0030} \\
ILM-IT-UI(unseen)         & {\bf 0.0717} & {\bf 0.0481} & {\bf 0.1086} & {\bf 0.0600}& {\rm 0.0213} & {\rm 0.0162} & {\rm 0.0269} & {\rm 0.0181} & {\rm 0.0038} & {\rm 0.0024} & {\bf 0.0062} & {\bf 0.0032} \\
\bottomrule
  \end{tabular}
  }
\end{table}

\begin{table}[t!]
  \caption{Effects of phase 1 item-item and user-item contrastive losses on OpenP5 phase 1 final train and eval item-grounded text generation losses.}
  \label{openp5_phase1_loss_values}
  \centering
  \begin{tabular}{ccccccc}
\toprule
\multirow{2}{*}{Methods} & \multicolumn{2}{c}{ML1M} & \multicolumn{2}{c}{Beauty} & \multicolumn{2}{c}{Clothing} \\
\cmidrule(r){2-3} \cmidrule(r){4-5} \cmidrule(r){6-7}
& Train & Eval & Train & Eval & Train & Eval \\
\midrule
ILM-IT              & {\rm 0.0000} & {\rm 4.1699} & {\rm 1.0441} & {\rm 4.2643} & {\rm 0.2114} & {\rm 2.0530} \\
ILM-IT-II           & {\rm 0.1552} & {\rm 3.8675} & {\rm 2.0232} & {\rm 3.2567} & {\rm 0.4594} & {\rm 1.7338} \\
ILM-IT-UI           & {\rm 0.0089} & {\rm 4.0663} & {\rm 2.3420} & {\rm 3.3724} & {\rm 0.5498} & {\rm 1.6358} \\
\bottomrule
  \end{tabular}
\end{table}

\subsection{Effects of Number of Query Tokens}
\begin{figure}[t]
  \centering
  \includegraphics[width=\textwidth]{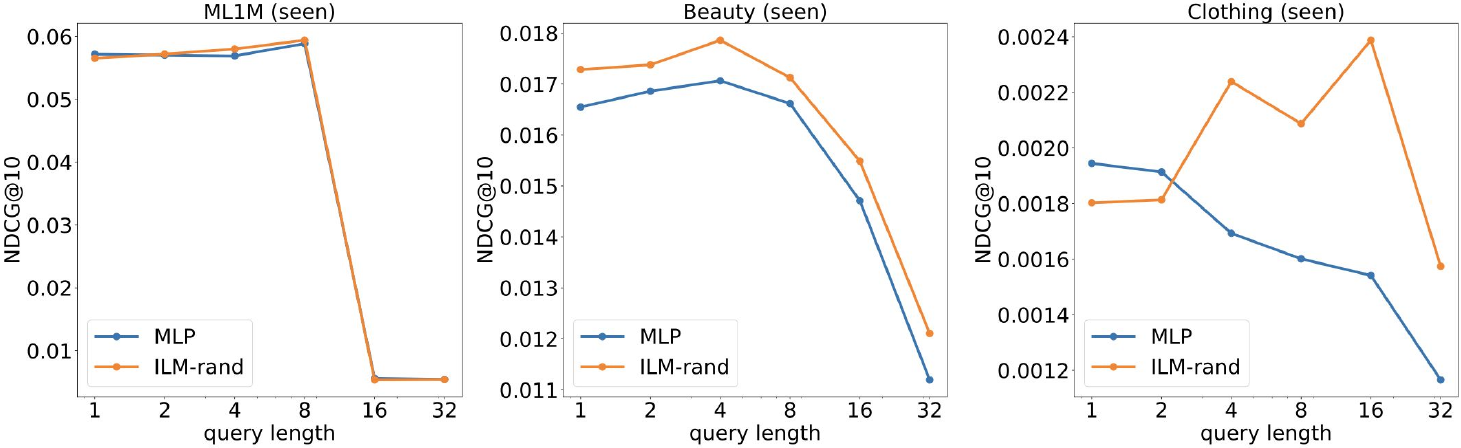}
  \caption{Effects of Number of Query Tokens.}
  \label{num_tokens_results}
\end{figure}

Another key aspect of our ILM approach is we used multiple learned queries to generate multiple embeddings in Q-Former output as item representation to feed into LLM. Existing methods~\cite{tennenholtz2024demystifying, zhang2023collm} typically use one embedding as the item-representation to feed into LLM.  We show ILM results using different numbers of queries tokens in Figure~\ref{num_tokens_results}. In order to better understand the gains of our approach, we also use the MLP approach to project the input embedding into a same number of embeddings. For both approaches, as the number of query tokens increases, the performance first increases then decreases. For most of the query lengths, our method outperforms the MLP approach.

\section{Conclusion}
\label{sec_conclusion}
We propose ILM, a novel approach for incorporating collaborative filtering knowledge into a frozen LLM for conversational recommendation tasks.  Specifically, we use a Q-Former item encoder to produce item-language aligned representations given item collaborative filtering embeddings, then interleave those representations with text token embeddings and feed into a frozen LLM. We propose a two-phase training paradigm. In Phase~1, we pretrain the Q-Former encoder using item-text representation learning objectives and a novel item-item representation learning objective. The item-item representation learning plays a regularization role to prevent the Q-Former from overfitting on item-text data if they are scarce. At the same time, it also directly encodes item-item similarity information in the Q-Former output representations. In Phase~2, we train ILM on conversational recommendation tasks by freezing LLM backbone parameters. This preserves the original pretrained LLM's language abilities, and reduces the privacy risk when finetuning LLM on user behavioral data. We conduct extensive benchmarks across various conversational recommendation datasets and tasks. \vvcomment{v- Added this line to make the claim clear.} We show the performance improvement due to item-item and user-item interaction data, demonstrating how this technique incorporates traditional CF signals. We demonstrate the value of this technique for both recommendation tasks and for language generation tasks for conversational recommendation.  The results show that our model can achieve SOTA performance consistently comparing existing methods of integrating collaborative filtering knowledge into a frozen LLM.

\newpage
\bibliography{neurips_2024}
\bibliographystyle{abbrv}

\newpage
\appendix

\section{Preserving Pretrained Ability}
In conversational recommendation, often the user and the system will conduct multiple turns of conversations, and for the system to achieve a certain goal, tool use may be employed \cite{feng2023large,gao2023chatrec,liu-etal-2023-conversational,fang2024multiagent}. There is often no constraint on the topics of the conversations, so the pretrained abilities of the LLM could be important. For example, the user may ask the LLM to perform certain operations such as adding or removing certain items from the recommended items, or conducting certain filtering based on a criteria. Those operations require certain reasoning ability from LLM pretraining. For another example, to use LLMs as an automatic agents \cite{agents_review,generative_agents} for conversational recommendation, they may require certain broad knowledge to perform tool use \cite{toolformer} to achieve a task. If the LLM is later fully finetuned only using the task specific data, it is likely that those pretrained abilities will be lost or hidden. Our ILM approach uses a frozen LLM, when the inputs don't contain items, the behavior of the model will be exactly the same as the original LLM. This means all pretrained knowledge can be preserved, which is crucial for multi-turn conversations and tool use in automatic agents.

\begin{table}
  \caption{The mean and standard error of the ILM metrics for ELM 24 tasks (computed using 3 runs with different random seeds)}
  \label{elm_mean_sem}
  \centering
  \begin{tabular}{lll}
\toprule
Tasks & SC(\%) & Log pplx \\
\midrule
summary              & 0.7318 $\pm$ 0.00359   & 0.6250 $\pm$ 0.00134    \\
positive review      & 0.7832 $\pm$ 0.00246   & 0.5711 $\pm$ 0.00211    \\
neutral review       & 0.7680 $\pm$ 0.00584   & 0.5759 $\pm$ 0.00345    \\
five pos char.       & 0.8631 $\pm$ 0.00320   & 0.5960 $\pm$ 0.00556    \\
five neg char.       & 0.9225 $\pm$ 0.00017   & 0.4029 $\pm$ 0.00845    \\
long description     & 0.7258 $\pm$ 0.00411   & 0.6622 $\pm$ 0.00130    \\
funnier              & 0.6891 $\pm$ 0.00358   & 0.5394 $\pm$ 0.00085    \\
sadder               & 0.7181 $\pm$ 0.00267   & 0.5152 $\pm$ 0.00112    \\
scarier              & 0.7241 $\pm$ 0.00261   & 0.5251 $\pm$ 0.00104    \\
improve              & 0.7759 $\pm$ 0.00293   & 0.5123 $\pm$ 0.00264    \\
movie to viewer      & 0.8008 $\pm$ 0.00313   & 0.6005 $\pm$ 0.00425    \\
pitch                & 0.8370 $\pm$ 0.00341   & 0.5181 $\pm$ 0.00198    \\
criticize            & 0.7894 $\pm$ 0.00356   & 0.5504 $\pm$ 0.00406    \\
convince1            & 0.7881 $\pm$ 0.00353   & 0.6176 $\pm$ 0.00353    \\
convince2            & 0.7957 $\pm$ 0.00523   & 0.7795 $\pm$ 0.00533    \\
convince3            & 0.7949 $\pm$ 0.00421   & 0.8900 $\pm$ 0.00714    \\
dissuade1            & 0.7909 $\pm$ 0.00203   & 0.6076 $\pm$ 0.00381    \\
dissuade2            & 0.8386 $\pm$ 0.00167   & 0.7278 $\pm$ 0.00609    \\
\midrule
similarities         & 0.8422 $\pm$ 0.00585   & 0.3993 $\pm$ 0.00842    \\
interpolation        & 0.7300 $\pm$ 0.00363   & 0.5310 $\pm$ 0.00238    \\
why like nn          & 0.8030 $\pm$ 0.00367   & 0.6452 $\pm$ 0.00556    \\
diff than nn         & 0.8891 $\pm$ 0.00198   & 0.5111 $\pm$ 0.00941    \\
common with nn       & 0.8433 $\pm$ 0.00182   & 0.5393 $\pm$ 0.00817    \\
\midrule
all                  & 0.7939 $\pm$ 0.00317   & 0.5813 $\pm$ 0.00415    \\
\bottomrule
  \end{tabular}
\end{table}

\begin{table}
  \caption{The mean and standard error of the ILM metrics for OpenP5 datasets (computed using 3 runs with different random seeds)}
  \label{openp5_mean_sem}
  \centering
  \resizebox{\textwidth}{!}{
  \begin{tabular}{lllll}
\toprule
Datasets & \multicolumn{1}{c}{HR@5} & \multicolumn{1}{c}{NDCG@5} & \multicolumn{1}{c}{HR@10} & \multicolumn{1}{c}{NDCG@10} \\
\midrule
ml1m(seen)           & {\rm 0.0715 $\pm$ 0.00051  } & {\rm 0.0476 $\pm$ 0.00047  } & {\rm 0.1072 $\pm$ 0.00043 }  & {\rm 0.0591 $\pm$ 0.00020 }   \\
beauty(seen)         & {\rm 0.0210 $\pm$ 0.00018  } & {\rm 0.0160 $\pm$ 0.00020  } & {\rm 0.0265 $\pm$ 0.00036 }  & {\rm 0.0177 $\pm$ 0.00025 }   \\
clothing(seen)       & {\rm 0.0040 $\pm$ 0.000085 } & {\rm 0.0025 $\pm$ 0.000014 } & {\rm 0.0061 $\pm$ 0.00022 }  & {\rm 0.0032 $\pm$ 0.000051}   \\
\midrule
ml1m(unseen)         & {\rm 0.0720 $\pm$ 0.00044  } & {\rm 0.0478 $\pm$ 0.00052  } & {\rm 0.1069 $\pm$ 0.00095 }  & {\rm 0.0591 $\pm$ 0.00049  }  \\
beauty(unseen)       & {\rm 0.0214 $\pm$ 0.00019  } & {\rm 0.0162 $\pm$ 0.000090 } & {\rm 0.0266 $\pm$ 0.00020 }  & {\rm 0.0179 $\pm$ 0.000099 }  \\
clothing(unseen)     & {\rm 0.0038 $\pm$ 0.000037 } & {\rm 0.0025 $\pm$ 0.000041 } & {\rm 0.0059 $\pm$ 0.00015 }  & {\rm 0.0032 $\pm$ 0.000029 }  \\
\bottomrule
  \end{tabular}
 }  
\end{table}

\section{Hyperparameters}
\label{sec_hparams}
The hyper-parameters for phase 1 and phase 2 trainings on the ELM 24 tasks are presented in Table~\ref{elm_hparams}.

\begin{table}[h!]
  \caption{Hyper-parameters for ELM 24 tasks.}
  \label{elm_hparams}
  \centering
  \begin{tabular}{c|cc}
\toprule
\multirow{6}{*}{Phase 1} & Q-Former & 8 layers, 168M params \\
& batch size & 256 \\
& learning rate & $3 \times 10^{-5}$ \\
& learning rate schedule & cosine decay \\
& optimizer & AdaFactor \cite{shazeer2018adafactor} \\
& \# steps & 259K \\
& hardware & 16 Cloud V5 TPUs \\
\midrule
\multirow{6}{*}{Phase 2} & LLM backbone & PaLM 2-S \cite{palm2} \\
& batch size & 32 \\
& learning rate & $5 \times 10^{-4}$ \\
& learning rate schedule & linear decay \\
& optimizer & AdaFactor \cite{shazeer2018adafactor} \\
& \# steps & 100K \\
& hardware & 64 Cloud V5 TPUs \\
\bottomrule
  \end{tabular}
\end{table}

For OpenP5 benchmarks, we use a 8 layer transformer decoder as the LLM backbone. We pretrain the LLM backbone using text only OpenP5 data to enable generative retrieval. For the ML1M dataset, we pretrain for 100K steps. For the Beauty dataset, we pretrain for 20K steps. For the Clothing dataset, we pretrain for 10K steps. The hyper-parameters for phase 1 and phase 2 trainings on the OpenP5 tasks are presented in Table~\ref{openp5_hparams}.

\begin{table}[h!]
  \caption{Hyper-parameters for OpenP5 tasks.}
  \label{openp5_hparams}
  \centering
  \begin{tabular}{c|cc}
\toprule
\multirow{6}{*}{Phase 1} & Q-Former & 8 layers, 168M params \\
& batch size & 256 \\
& learning rate & $3 \times 10^{-5}$ \\
& learning rate schedule & cosine decay \\
& optimizer & AdaFactor \cite{shazeer2018adafactor} \\
& \# steps & 40K for ML1M, 10K for Beauty, 15K for Clothing \\
& hardware & 16 Cloud V5 TPUs \\
\midrule
\multirow{6}{*}{Phase 2} & LLM backbone & Transformer decoder 8 layers, 128M params \\
& batch size & 32 \\
& learning rate & $5 \times 10^{-4}$ \\
& learning rate schedule & linear decay \\
& optimizer & AdaFactor \cite{shazeer2018adafactor} \\
& \# steps & 50K for ML1M, 20K for Beauty, 20K for Clothing \\
& hardware & 64 Cloud V5 TPUs \\
\bottomrule
  \end{tabular}
\end{table}

\end{document}